\documentclass[twocolumn,showpacs,preprintnumbers,amsmath,amssymb,superscriptaddress,showkeys]{revtex4}

\usepackage{graphicx}
\usepackage{dcolumn}
\usepackage{bm}

\begin{document}

\preprint{}

\title{Optical design of reflectionless complex media by finite embedded coordinate transformations}

\author{Marco Rahm}
\email{marco.rahm@duke.edu}
\author{Steven A. Cummer}
\author{David Schurig}
\affiliation{%
Department of Electrical and Computer Engineering, Duke University,
Durham, NC 27708, USA
}%
\author{John B. Pendry}
\affiliation{ Department of Physics, The Blackett Laboratory,
Imperial College, London SW7 2AZ, UK
}%
\author{David R. Smith}
\affiliation{%
Department of Electrical and Computer Engineering, Duke University,
Durham, NC 27708, USA
}%
\affiliation{ Department of Physics, The Blackett Laboratory,
Imperial College, London SW7 2AZ, UK
}%

\date{\today}

\begin{abstract}
Transformation optics offers an unconventional approach to the
control of electromagnetic fields.  A transformation optical
structure is designed by first applying a form-invariant coordinate
transform to Maxwell's equations, in which part of free space is
distorted in some desired manner.  The coordinate transformation is
then applied to the permittivity and permeability tensors to yield
the specification for a complex medium with desired functionality.
The transformation optical structures proposed to date, such as
electromagnetic "invisibility" cloaks and concentrators, are
inherently reflectionless and leave the transmitted wave
undisturbed.  Here we expand the class of transformation optical
structures by introducing finite, embedded coordinate
transformations, which allow the electromagnetic waves to be steered
or focused.  We apply the method to the design of several devices,
including a parallel beam shifter and a beam splitter, both of which
exhibit unusual electromagnetic behavior as confirmed by 2D
full-wave simulations.  The devices are designed to be
reflectionless, in accordance with a straightforward topological
criterion.
\end{abstract}

\maketitle

Metamaterials offer an enormous degree of freedom for manipulating
electromagnetic fields, as independent and nearly arbitrary
gradients can be introduced in the components of the effective
permittivity and permeability tensors. In order to exploit such a
high degree of freedom a viable method for the well-aimed design of
complex materials is required. Pendry et al.\ reported a methodology
based on continuous form-invariant coordinate transformations of
Maxwell's equations which allows for the manipulation of
electromagnetic fields in a previously unknown and unconventional
fashion \cite{Pendry:2006}. This method was successfully applied for
the design and the experimental realization of an invisibility cloak
\cite{Schurig2:2006}. This work generated widespread interest
specifically in the prospects of electromagnetic cloaking, a topic
that has dominated much of the subsequent discussion
\cite{Greenleaf:2006,Leonhardt:2006,Milton:2006,Milton2:2006,
Cummer:2007,Cai:2007,Ruan:2007,Chen:2007,Wood:2007,Cummer:2006,Rahm:2007,
Kong:2007,Cai2:2007}. All the transformation-optical designs
reported in the literature so far have in common that the
electromagnetic properties of the incident waves are exclusively
changed within the restricted region of the transformation-optical
device. However, for continuous transformations the field
manipulation cannot be transferred to another medium or free space
and thus remains a local phenomenon.

In this letter, we report a generalized approach to the method of
form-invariant coordinate transformations of Maxwell's equations
based on finite embedded coordinate transformations. The use of
embedded transformations adds a significant amount of flexibility to
the transformation design of complex materials. For example, with
finite-embedded transformations, it is possible to transfer field
manipulations from the transformation-optical medium to a second
medium, eliminating the requirement that the transformation optical
structure be invisible to an observer. The finite-embedded
transformation thus significantly broadens the range of materials
that can be designed to include device-type structures capable of
focusing or steering electromagnetic waves.  Like transformation
optical devices, the finite-embedded transform structures can be
reflectionless under conditions that we describe below.

The mathematical formalism used for the calculation of the complex
material properties is similar to the one reported in
\cite{Schurig:2006,Rahm:2007}. Throughout this paper, we use the
same terminology as introduced in these references.

For a given coordinate transformation $x^{\alpha'}(x^{\alpha}) =
A^{\alpha'}_{\alpha} x^{\alpha}$ ($A^{\alpha'}_{\alpha}$: Jacobi
matrix, $\alpha=1 \ldots 3$), the electric permittivity
$\epsilon^{i' j'}$ and the magnetic permeability $\mu^{i' j'}$ of
the resulting material can be calculated by
\begin{eqnarray}
\epsilon^{i' j'} &=&
[det(A^{i'}_{i})]^{-1}A^{i'}_{i}A^{j'}_{j}\epsilon^{i j}
\label{eqn:epsilon} \\
\mu^{i' j'} &=& [det(A^{i'}_{i})]^{-1}A^{i'}_{i} A^{j'}_{j}\mu^{i
j}\label{eqn:mu}
\end{eqnarray}
where $det((A^{i'}_{i})$ denotes the determinant of the Jacobi
matrix. For all the transformations carried out in this letter, the
mathematical starting point is 3-dimensional euclidian space
expressed in cartesian coordinates with isotropic permittivities and
permeabilities $\epsilon^{i j} = \varepsilon_{0} \delta^{i j}$ and
$\mu^{i j} = \mu_{0} \delta^{i j}$ ($\delta^{i j}$: Kronecker
delta).

A possible coordinate transformation for the design of a parallel
beam shifter and a beam splitter consisting of a slab with thickness
$2d$ and height $2h$ can be expressed by
\begin{eqnarray}
 x'(x,y,z)& = & x \label{eqn:xtrafo}\\
 y'(x,y,z)& = & \Theta(h_1\!\!-\!|y|)\big[ y+ak_l(x,y) \big] \nonumber \\
 & +& \Theta(|y|\!\!-\!h_1)\big[ y+\gamma(y) k_l(x,y)(y-s_2(y) h) \big] \nonumber \\
\label{eqn:ytrafo} \\
 z'(x,y,z) & = & z \label{eqn:ztrafo}
\end{eqnarray}
with
\begin{equation}
\Theta: \xi \rightarrow \Theta(\xi) :=
\begin{cases}
  \begin{array}{cc} 1 & \xi> 0 \\
                    1/2 & \xi=0 \\
                    0 & \xi < 0
  \end{array}
\end{cases}
\end{equation}
\begin{equation}
k_l: (\eta,\kappa) \rightarrow k_l(\eta,\kappa) :=
s_p(\kappa)(\eta+d)^l
\end{equation}
\begin{equation}
s_p: \xi \rightarrow s_p(\xi) :=
\begin{cases}
  \begin{array}{ccc} 1 & & p=1 \\
  \begin{cases}                   +1  \hspace{0.5in} \xi\geq 0\\
                                 -1 \hspace{0.5in} \xi<0
  \end{cases} & & p=2
  \end{array}
\end{cases}
\end{equation}
\begin{equation}
\gamma: \vartheta \rightarrow \gamma(\vartheta) :=
\frac{a}{s_2(\vartheta) (h_1-h)}
\end{equation}
where $2h_1$ is the maximum allowed width of the incoming beam, $a$
determines the shift amount and $l=1\ldots n$ is the order of the
nonlinearity of the transformation.

The transformation equations are defined for $(|x|\leq d)$,
$(|y|\leq h)$ and $(|z|< \infty)$. For the case $p=1$, equations
(\ref{eqn:xtrafo})-(\ref{eqn:ztrafo}) describe a parallel beam
shifter whereas for $p=2$ the equations refer to a beam splitter.
The Jacobi matrix of the transformation and its determinant are
\begin{equation}
A^{i'}_{i}  =  \left(
\begin{array}{ccc} 1 & 0 & 0 \\
a_{21} &
a_{22} & 0 \\
0 & 0 & 1
\end{array} \right)
\end{equation}
\begin{equation}
det(A^{i'}_{i}) = a_{22}
\end{equation}
with
\begin{eqnarray}
a_{21} & = & \Theta\big( h_1\!\!-\!|f'_1|\big)\big[ l a k'_{l-1}
\big] \nonumber \\
& + &  \Theta\big(|f'_2|\!\!-\!h_1\big) \big[l \gamma'
k'_{l-1}\big(f'_2-s_2(y')h\big)\big] \\
a_{22} & = & \Theta\big(h_1\!\!-\!|f'_1|\big) +
\Theta\big(|f'_2|\!\!-\!h_1\big) \big[ 1+ \gamma' k'_l \big]
\end{eqnarray}
where
\begin{eqnarray}
f'_1: (x',y')\rightarrow f(x',y') &:=& y'-a k'_l\\
f'_2: (x',y')\rightarrow f(x',y') &:=& \frac{y'+ \gamma' k'_l
s_2(y') h}{1+\gamma' k'_l}
\end{eqnarray}
$k'_l:=k_l(x',y')$ and $\gamma':=\gamma(y')$.

By equations (\ref{eqn:epsilon})-(\ref{eqn:mu}) it is
straightforward to calculate the tensors of the transformed relative
electric permittivity $\epsilon_r=\epsilon/\varepsilon_0$ and the
relative magnetic permeability $\mu_r=\mu/\mu_0$, which in the
material representation are obtained as
\begin{equation}
\epsilon_r^{ij} = \mu_r^{ij}  = \frac{1}{a_{22}} g^{ij}
\label{eqn:MatProperties}
\end{equation}
where
\begin{equation}
g^{ij} = \left(
\begin{array}{ccc} 1 & a_{21} & 0 \\
a_{21} &
a_{21}^2+ a_{22}^2 & 0 \\
0 & 0 & 1
\end{array} \right)
\end{equation}
is the metric tensor of the coordinate transformation.
\begin{figure}
\centering
\includegraphics[width = 3in]{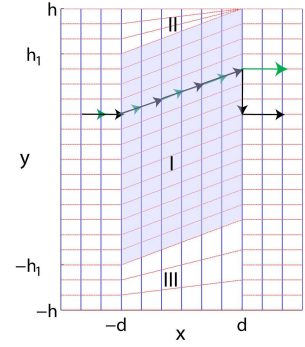}
\caption{\label{fig:Fig1} Linear spatial coordinate transformation
of a parallel beam shifter.}
\end{figure}
At this point it should be mentioned, that only the domain with
$\Theta(h_1\!\!-\!|f'_1|)\equiv1$ has to be considered in the
material implementation which simplifies the mathematical
expressions.

A linear coordinate transformation (l=1) for the design of a
parallel beam shifter is illustrated in Fig.\ \ref{fig:Fig1}. The
space within the grey shadowed region I
($\Theta(h_1\!\!-\!|f'_1|)\equiv1$) is tilted at an angle $\phi =
\arctan(a)$ with respect to free space, which inherently results in
a compression of space in region II and a dilution of space in
region III. Considering the boundary between region I and free
space, the coordinate systems are seen to be discontinuous at the
interface $x=d$. At this point, it is important to explain the
difference between ''embedded transformations'' and ''discontinuous
transformations''. Interpreting the transformation as discontinuous,
the boundary must be taken into account and the transformation of
the y-coordinate at the transition from region I and free space must
read
\begin{equation}
y'(x,y,z) =  \Theta(d-x) \big[y+ak_l(x,y) \big]+\Theta(x-d)y
\end{equation}
so that $a_{21}|_{(x=d)}\propto \delta(d-x)$. As in the ray
approximation the y-coordinate lines in fig.\ \ref{fig:Fig1}
represent the direction of the power flow for a beam propagating
from $x=-\infty$ to $\infty$. The inclusion of the delta
distribution in the material parameters would result in the
trajectory indicated by the black arrows. The incoming wave would be
shifted in the y-direction and abruptly be forced back on its old
path at the boundary at $x=d$, thereby rendering the entire beam
shifting section invisible to an observer. However, as can be seen
from equation (\ref{eqn:ytrafo}), the boundary is not included in
the calculation of the material properties for the beam shifter and
beam splitter. The coordinate transformation is carried out locally
for the transformation-optical medium and then embedded into free
space which results in the trajectory indicated by the green arrows.
The beam is shifted in the y-direction and maintains its lateral
shift after exiting the transformation-optical medium. This method
is similar to the ''embedded Green function'' approach in the
calculation of electron transport through interfaces
\cite{Inglesfield:1981}, so that we call it an ''embedded coordinate
transformation''.

At this point, the question arises as to which conditions must hold
for the embedded transformation in order to design a reflectionless
optical device. We found as a necessary -- and in our investigated
cases also sufficient -- topological condition that the metric in
and normal to the interface between the transformation-optical
medium and the non-transformed medium (in this case free space) must
be continuous to the surrounding space. In the case of the beam
shifter this means that the distances as measured along the x-, y-
and z-axis in the transformation-optical medium and free space must
be equal along the boundary ($x=d$). As can be clearly seen from
fig.\ \ref{fig:Fig1}, this condition is fulfilled by the embedded
coordinate transformation of the beam shifter within the green
shadowed region I. It should be mentioned, that the material
properties in the domains II and III can be arbitrarily chosen as no
fields penetrate into these regions.
\begin{figure}
\centering
\includegraphics[width = 3in]{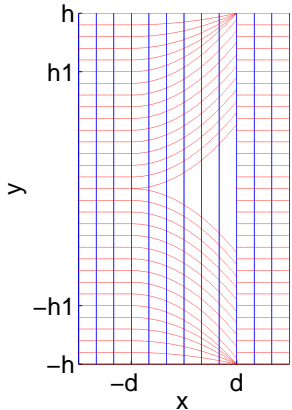}
\caption{\label{fig:Fig2} Nonlinear spatial coordinate
transformation of second order for a beam divider.}
\end{figure}

As a second, more sophisticated example a beam splitter is presented
for the case of a nonlinear transformation of second order. The
material properties are described by equation
(\ref{eqn:MatProperties}) with ($p=2$) and ($l=2)$. This specific
coordinate transformation is illustrated in fig.\ \ref{fig:Fig2}.
The underlying metric describes the gradual opening of a
wedge-shaped slit in the y-direction. The metric in the x-direction
is not affected by the transformation. Similar to the parallel beam
shifter, the beam splitter obeys the topological condition in order
to operate without reflection.

To confirm our findings, 2D full-wave simulations were carried out
to adequately predict the electromagnetic behavior of waves
impinging on a beam shifter and a beam divider, respectively. The
calculation domain was bounded by perfectly matched layers. The
polarization of the plane waves was chosen to be perpendicular to
the x-y plane.
\begin{figure*}
\centering
\includegraphics[width = 4.5in]{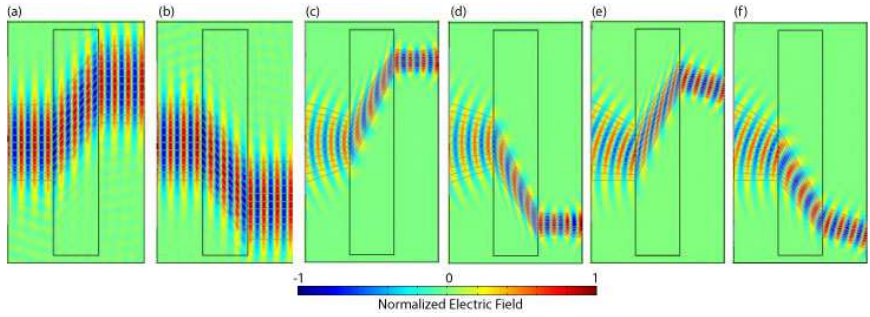}
\caption{\label{fig:Fig3} Electric field distribution (color map)
and power flow lines (grey) of a parallel beam shifter for
diffracting plane waves with shift parameters (a) $a=1.8$, (b) -1.8,
for a convergent beam under perpendicular incidence with (c) $a=2$,
(d) $=-2$ and for oblique incidence with (e) $a=2$ and (f)
$a=-1.2$.}
\end{figure*}

Fig.\ \ref{fig:Fig3} depicts the spatial distribution of the real
part of the transverse-electric phasor (color map) and the direction
of the power flow (grey lines) of propagating waves across a
parallel beam shifter at perpendicular (fig.\ \ref{fig:Fig3}a-d) and
oblique incidence (fig.\ \ref{fig:Fig3}e-f). The curvature of the
incoming wave fronts was freely chosen to be plane (a-b) or
convergent (c-f). As can be seen from fig.\ \ref{fig:Fig3}a-b, the
beam shifter translates the incoming plane wave in the y-direction
perpendicular to the propagation x-direction without altering the
angle of the wave fronts. In contrast, the direction of the power
flow changes by an angle $\phi=\\arctan (a)$ ($a$: shift parameter)
with respect to the power flow of the incoming plane wave. After
propagation through the complex transformation-optical medium, the
wave fronts and the power flow possess the same direction as the
incoming beam, however the position of the wave is offset in the
y-direction. The shift parameter $a$ was arbitrarily chosen to be
1.8 (fig.\ \ref{fig:Fig3}a) and -1.8 (fig.\ \ref{fig:Fig3}b).

A similar behavior can be observed for waves with wave fronts of
arbitrary curvature, as for example for convergent waves (fig.\
\ref{fig:Fig3}c-f). In this case, the focus of the beam can be
shifted within a plane parallel to the y-axis by variation of the
shift parameter $a$. As is obvious from fig.\ \ref{fig:Fig3}e-f, the
same behavior applies for incoming waves at oblique incidence. The
beam solely experiences a translation in the y-direction whereas the
x-position of the focus remains unchanged. In all cases, the
realized transformation-optical parallel beam shifter proves to
operate without reflection confirming our metric criterion used for
the design. The presented parallel beam shifter could play a crucial
role in connection with tunable, reconfigurable metamaterials as it
would allow scanning of a beam focus along a flat surface without
changing the plane of the focus and without introducing a beam tilt
or aberrations. These properties become even more significant for
applications where short working distances are used between scanner
and object.
\begin{figure*}
\centering
\includegraphics[width = 4in]{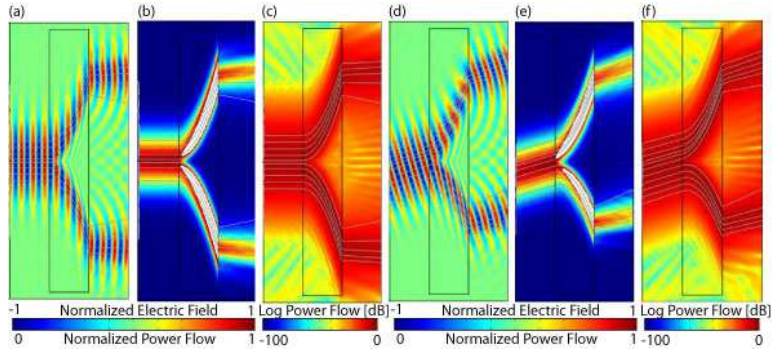}
\caption{\label{fig:Fig4} Electric field distribution (color map),
power flow lines (grey) (a+d) and power flow on a linear (b+e) and
logarithmic scale (c+f) of a beam splitter for diffracting plane
waves with shift parameters (a-c) $|a|=15$ for perpendicular
incident waves and (d-f) $|a| = 12$ for oblique incidence.}
\end{figure*}

Similar simulations were carried out for the transformation-optical
beam splitter. Fig.\ \ref{fig:Fig4}a shows the electric field
distribution and the power flow lines for waves at perpendicular
incidence. The beam splitter shifts one half of the wave in the
(+y)-direction and the second half in the (-y)-direction, thus
splitting the wave at the mid-point. The split waves are not
perfectly parallel at the exit plane of the device due to
diffraction of the incoming wave of finite width. As can be seen,
there exists a small fraction of scattered fields within the split
region which can be explained in terms of diffraction and scattering
which is out of the scope of this letter. The beam splitter was
found to operate without reflection in agreement with the metric
criterion.

Fig.\ \ref{fig:Fig4}b displays the normalized power flow inside and
outside the device. In order to enhance the contrast in the
visualization of the power distribution at the beam splitter output,
the color scale is saturated inside the beam splitter medium. As
obvious from the transformation (fig.\ \ref{fig:Fig2}), the power
density inside the transformation-optical medium is expected to be
higher than outside the material, which is indicated by the density
of the grid lines in fig.\ \ref{fig:Fig2} and confirmed by the
simulations. The power flow density abruptly decreases at the output
facet of the beam splitter. For clarification, fig.\ \ref{fig:Fig4}c
shows the power flow on a logarithmic scale. By integration of the
power density inside the gap region between the beams and the power
density inside either the upper or lower arm of the split beams a
power ratio of 10:1 was calculated. The scattered waves in the gap
carried about 4\% of the total power.

In fig.\ \ref{fig:Fig4}d-f the spatial distribution of the
transverse electric field (d) and the power flow on a linear (e) and
a logarithmic scale (f) are shown for an obliquely incident wave on
the beam splitter. Again, the beam is clearly divided into two beams
with a small fraction of diffracted and scattered fields inside the
gap between the split beams. As for perpendicular incidence, no
reflection was observed in the simulation at both the input and
output facet of the beam splitter. Similar to the result for
perpendicular incidence, the propagation directions of the outgoing
waves are not parallel. In addition, the angles of the central wave
vectors of the split beams with refer to the central wave vector of
the incident beam are not equal.

In conclusion, the concept of embedded coordinate transformations
significantly expands the idea of the transformation-optical design
of metamaterials which itself was restricted to continuous
coordinate transformations so far. In contrast to continuous
transformations, the expansion to embedded transformations allows
for the first time to non-reversibly change the properties of
electromagnetic waves in transformation media and to transmit the
changed electromagnetic properties to free space or to a different
medium in general. In order to design the medium as reflectionless,
a new topological criterion for the embedded transformations was
found, which imposes constraints to the metric of the spaces at the
interface between the transformation-optical medium and the
surrounding space. This metric criterion was applied in the
conception of a parallel beam shifter and a beam splitter and
confirmed in 2D full wave simulations. Both devices showed an
extraordinary electromagnetic behavior which is not achievable by
conventional materials. Both examples clearly state the significance
of embedded coordinate transformations for the design of new
electromagnetic elements with tunable, unconventional optical
properties.

D.\ Schurig wishes to acknowledge support from the IC Postdoctoral
Research Fellowship program.

\end{document}